\documentstyle[aas2pp4,psfig]{article}

\newcommand\mdot   {\hbox {${\dot M}$}}
\newcommand\sz     {S$_{\rm z}$}

\newcommand\pp     {$\pm$}

\newcommand\micros  {$\mu$s}

\righthead{KiloHertz Quasi-Periodic Oscillations in GX~17$+$2}
\slugcomment{Submitted to ApJ Letters, September 1997}

\begin{document}

\title{Discovery of KiloHertz Quasi-Periodic Oscillations in GX\,17$+$2}

\author{Rudy Wijnands\altaffilmark{1},
        Jeroen Homan\altaffilmark{1},
        Michiel van der Klis\altaffilmark{1},
	Mariano M\'endez\altaffilmark{1,2},
	Erik Kuulkers\altaffilmark{3}.
	Jan van Paradijs\altaffilmark{1,4},
	Walter H. G. Lewin\altaffilmark{5},
	Frederick K. Lamb\altaffilmark{6},
	Dimitrios Psaltis\altaffilmark{6},
        Brian Vaughan\altaffilmark{7}
        }

\altaffiltext{1}{Astronomical Institute ``Anton Pannekoek'',
University of Amsterdam, and Center for High Energy Astrophysics,
Kruislaan 403, NL-1098 SJ Amsterdam, The Netherlands;
rudy@astro.uva.nl, homan@astro.uva.nl, michiel@astro.uva.nl,
mariano@astro.uva.nl, jvp@astro.uva.nl}

\altaffiltext{2}{Facultad de Ciencias Astron\'omicas y
       Geof\'{\i}sicas, Universidad Nacional de La Plata, Paseo del
       Bosque S/N, 1900 La Plata, Argentina}

\altaffiltext{3}{Astrophysics, University of Oxford,
Nuclear and Astrophysics Laboratory, Keble Road, Oxford OX1 3RH,
United Kingdom; e.kuulkers1@physics.oxford.ac.uk}

\altaffiltext{4}{Departments of Physics,
University of Alabama at Huntsville, Huntsville, AL 35899}

\altaffiltext{5}{Department of Physics and Center for Space Research,
Massachusetts Institute of Technology, Cambridge, MA 02139;
lewin@space.mit.edu}

\altaffiltext{6}{Departments of Physics and Astronomy,
University of Illinois at Urbana-Champaign, Urbana, IL 61801;
f-lamb@uiuc.edu, demetris@astro.uiuc.edu}

\altaffiltext{7}{Space Radiation Laboratory, California
Institute of Technology, 220-47 Downs, Pasadena, CA 91125;
brian@thor.srl.caltech.edu}

\begin{abstract}

We observed the low-mass X-ray binary and Z source GX~17$+$2 with the
Rossi X-ray Timing Explorer during 6--8 Feb 1997, 1--4 Apr 1997, and
26--27 Jul 1997. The X-ray color-color diagram shows a clear Z track.
Two simultaneous kHz quasi-periodic oscillations (QPOs) are present in
each observation, whose frequencies are well correlated with the
position of the source on the Z-track. At the left end of the
horizontal branch (HB) only the higher frequency peak is observed, at
645$\pm$9 Hz, with an rms amplitude of 5.7$\pm$0.5\%, and a FWHM of
183$\pm$35 Hz. When the source moves down the Z track to the upper
normal branch the frequency of the kHz QPO increases to 1087\pp12 Hz,
and the rms amplitude and FWHM decrease by a factor 2.  Further down
the Z the QPO becomes undetectable, with rms upper limits of typically
2.0\%. Halfway down the HB a second QPO appears in the power spectra
with a frequency of 480\pp23 Hz. The frequency of this QPO also
increases when the source moves along the Z track, up to 781\pp11 Hz
halfway down the normal branch, while the rms amplitude and FWHM stay
approximately constant at 2.5\% and 70 Hz.  The QPO frequency
difference is constant at 293.5\pp7.5 Hz.  Simultaneously with the kHz
QPOs we detect HB QPOs (HBOs). The simultaneous presence of
HBOs and kHz QPOs excludes the magnetospheric beat-frequency model as
the explanation for at least one of these two phenomena.  

\end{abstract}


\section{Introduction \label{intro}}

GX 17$+$2 is a bright low-mass X-ray binary and Z source (Hasinger \&
van der Klis 1989).  Along the Z track traced out in the X-ray
color-color diagram (CD) the mass accretion rate (\mdot) is thought to
increase from the horizontal branch (HB), via the normal branch (NB),
to the flaring branch (FB).  On the HB and upper NB 18--60 Hz
quasi-periodic oscillations (QPOs) are present (called HBOs), and on
the lower NB and FB 5--20 Hz QPOs (N/FBOs).  Atoll sources have no Z
track in the CD (Hasinger \& van der Klis 1989), and do not exhibit
HBO or N/FBO\footnote{Although in several of them broad QPO-like peaks
are found with frequency around 20 Hz (e.g. Hasinger \& van der Klis
1989; Wijnands \& van der Klis 1997)}. However, many show kHz QPOs
between 300 and 1200 Hz (see van der Klis 1997 for a recent
review). Usually two simultaneous kHz QPOs (hereafter: twin peaks) are
seen, whose frequencies increase with \mdot, while their difference
(the twin peak separation) remains constant. In X-ray bursts
oscillations are seen with a frequency once or twice the twin peak
separation. This strongly suggest a model in which the burst
oscillation frequency is the spin frequency (or twice that), the
higher kHz QPO frequency some inner disk frequency, and the lower kHz
QPO frequency their beat.

So far, two Z sources have shown kHz QPOs: Sco X-1 with frequencies in
the range 550--1100 Hz (van der Klis et al. 1996a; 1997b), and GX
5$-$1 (325--896 Hz; van der Klis et al. 1996b).  Both sources showed
twin peaks.  In GX 5$-$1 the QPO frequencies increased from the left
end of the HB to the upper NB (van der Klis et al. 1996b), in Sco X-1
from the upper NB to the FB (van der Klis et al. 1996a; 1997b).  In
view of the similarities between the kHz QPOs in Z and atoll sources,
both are most likely caused by the same mechanism. The only difference
so far is that in atoll sources the peak separation is constant,
whereas in Sco X-1 it decreases when the QPO frequencies increase.  In
this Letter, we report the discovery of twin kHz QPOs in the Z source
GX~17$+$2 (see also van der Klis et al. 1997a).

\section{Observations and Analysis  \label{observations}}

We observed GX 17$+$2 with the PCA onboard the Rossi X-ray Timing
Explorer on 6--8 Feb, 1--4 Apr, and 26--27 Jul 1997. A total of 120.5
ksec of good data were obtained. During Feb 8 (1.1 ksec) and a small
part of Feb 7 (7.8 ksec) and Jul 27 (4.8 ksec) only 4 of the 5
detectors were on. The source then covered parts of the Z also
well-covered with 5 detectors.  As adding the 4-detector data did not
improve the results significantly, we only used the 5-detector data to
have the most accurate CD possible.  Data were collected with 16 s
time resolution in 129 photon energy bands (energy range 2--60 keV),
and simultaneously with 122 \micros\, time resolution in 4 bands: 2--5
keV, 5--6.4 keV, 6.4--8.6 keV, and 8.6--60 keV.

We made power density spectra from the 122 \micros\, data, using 16 s
data segments.  To measure the kHz QPOs we fitted the 96--4096 Hz
power spectra with a function described by a constant, one or two
Lorentzian peaks, and a broad sinusoid to represent the dead-time
modified Poisson noise (Zhang et al. 1995; Zhang 1995).  The Very
Large Event window (van der Klis et al. 1997b) was set to 55 \micros,
so that its effect on the Poisson noise was small enough that it could
be absorbed by the broad sinusoid.  To measure the HBOs we fitted the
8--256 Hz power spectra with a constant, one or two Lorentzian peaks,
and a power law representing the continuum. Differential dead time
(van der Klis 1989) was negligible.  We determined errors using
$\Delta\chi^2 = 1.0$ and upper limits using $\Delta\chi^2 = 2.71$,
corresponding to a 95\% confidence level. Upper limits on the kHz QPOs
were determined using a fixed FWHM (100 Hz).  In the presence of one
kHz peak the upper limit on the other was determined by fixing its
frequency at the frequency of the detected QPO plus or minus the mean
peak separation (depending on whether the lower or higher frequency
QPO was detected, respectively).  Upper limits on the HBO second
harmonic were determined using a fixed FWHM (10 Hz) and a frequency
twice that of the first harmonic.

In the CD and the hardness-intensity diagram (HID) we used for the
soft color the log of the 3.5--6.4 keV/2.0--3.5 keV count rate ratio,
for the hard color the log of the 9.7--16.0 keV/6.4--9.7 keV ratio,
and as intensity the log of the count rates in the 2.0--16.0 keV band.

We used the \sz\, parameterization (Wijnands et al. 1997a, and
references therein) for measuring the position along the Z.  The HB/NB
vertex is at \sz=1.0, the NB/FB vertex at \sz=2.0.  By using
logarithmic values for the colors, \sz\, does not depend on the values
of the colors but only on their variations (Wijnands et al. 1997a).
We selected the power spectra according to \sz, and determined the
average \sz\, for each average power spectrum. The \sz\, error bars
represent the standard deviation of the \sz\, distributions.

\section{Results \label{results}}

\begin{figure}[t]
\begin{center}
\begin{tabular}{c}
\psfig{figure=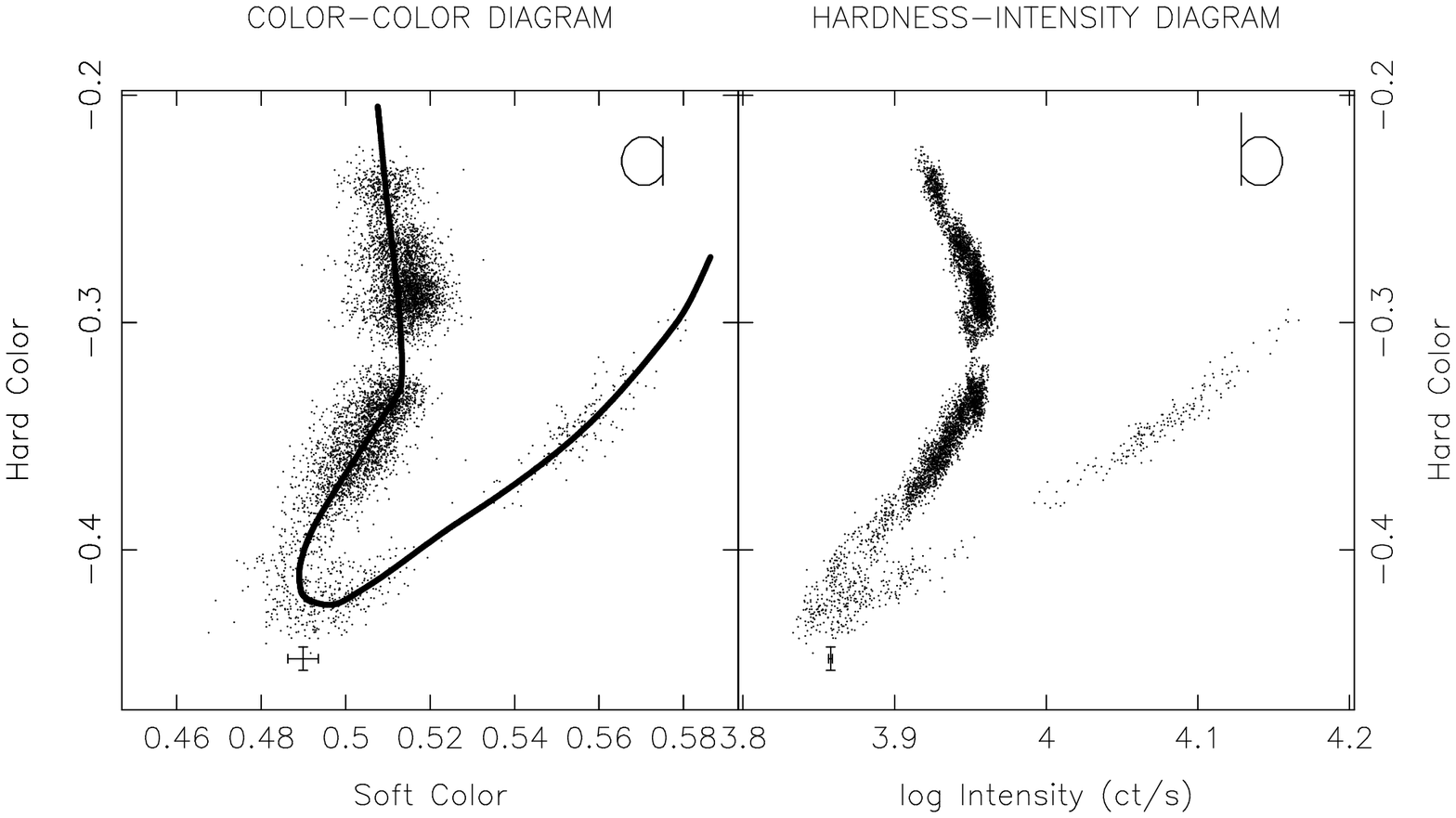,width=18cm}
\end{tabular}
\caption{Color-color diagram ({\it a}) and
hardness-intensity diagram ({\it b}) of GX 17$+$2. The soft color is
the logarithm of the ratio of the count rates between 3.5--6.4 keV and
2.0--3.5 keV; the hard color is the logarithm of the ratio of the
count rates between 9.7--16.0 keV and 6.4--9.7 keV; the intensity is
the logarithm of the count rate in the photon energy range 2.0--16.0
keV. The background, which is almost negligible (50 ct/s in the energy
range 2.0--16.0 keV), was subtracted, but the intensity was not
dead-time corrected. All points are 16 s averages.  The thick line in
the CD is the spline which was used to determine \sz. Typical error
bars are shown at one of the points. \label{cd}}
\end{center}
\end{figure}

In April the source traced out a full Z in the CD and HID
(Fig.~\ref{cd}). In February the HB/NB vertex and the upper NB were
covered, and in July the HB.  The Feb data fall right on top of the
Apr data, indicating that the Z did not move appreciably between
February and April. The Jul data have higher count rate and higher
soft color. Such shifts have been observed before in GX 17$+$2 by
Kuulkers et al. (1997), and also in other Z sources (Cyg X-2:
Hasinger et al. 1990; GX 5$-$1: Kuulkers et al. 1994).  In all
observations we detected kHz QPOs in the 8.6--60 keV band.  In the
6.4--8.6 keV band the QPOs were not, or marginally,
detected. Combining both bands the significance of the QPOs increased,
although the rms amplitude decreased a little. Therefore, we used the
6.4--60 keV range for the analysis of the QPOs.

\begin{figure}[t]
\begin{center}
\begin{tabular}{c}
\psfig{figure=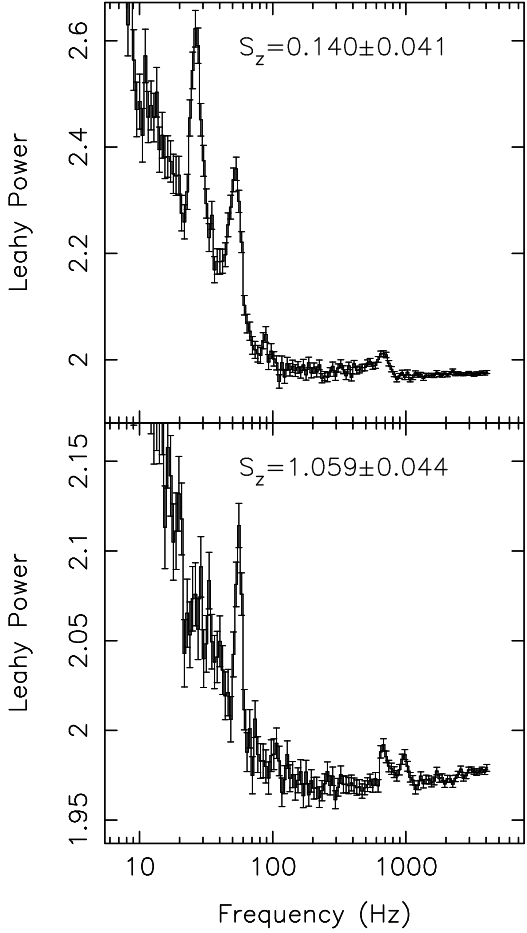,width=7cm}
\end{tabular}
\caption{Typical Leahy normalized power spectra
in the energy range 6.4--60 keV. In the top panel, for
\sz=0.140\pp0.041, clearly the HBO, its second harmonic, and the
higher frequency QPO at 675 Hz are visible. In the bottom panel, for
\sz=1.059\pp0.044, the HBO fundamental, the lower frequency QPO (680
Hz) and the higher frequency QPO (986 Hz) are visible. The upward
trend of the power at kHz frequencies, best seen in the bottom panel,
is due to instrumental dead time.
\label{powerspectrum}}
\end{center}
\end{figure}

We detected the kHz QPO only for \sz$<1.5$.  Typical power spectra are
shown in Fig.~\ref{powerspectrum}. Note the simultaneous presence of
the HBO and its harmonic (upper frame), and the kHz QPOs.  The
properties of the QPOs as a function of \sz\, are shown in
Fig.~\ref{QPOs}. The higher frequency QPOs increase in frequency from
645\pp9 Hz at \sz=0.07\pp0.05 to 1087\pp12 Hz at \sz=1.44\pp0.05
(Fig.\,\ref{QPOs}a), while their rms amplitude and FWHM decrease from
5.7\pp0.5\% to 1.5\pp0.4\% (Fig\,\ref{QPOs}c), and from 183\pp35 Hz to
21\pp20 Hz. Above \sz=1.5 the higher frequency QPO is undetectable
(upper limits typically 2.5\% rms).  For \sz$<$0.5 the lower frequency
QPO is not detected, with upper limits near 2.5\% rms.  From
\sz=0.56\pp0.04 to \sz=1.34\pp0.05 the lower frequency QPO is detected
at a frequency increasing from 480\pp23 Hz to 781\pp11 Hz
(Fig.\,\ref{QPOs}a).  The rms amplitude (Fig.\,\ref{QPOs}e) and FWHM
of the lower frequency QPO are about constant between 2--3\% and
50--150 Hz. When both QPOs are detected the frequency difference is
consistent with being constant at 293.5\pp7.5 Hz.

The QPOs were strongest at high energy.  We combined all data with
\sz$<$0.5 (\sz=0.23\pp0.14), and with \sz\, between 1.0 and 1.3
(\sz=1.15\pp0.10) to detect the QPOs at lower energy.  When \sz$>$1.0
both QPOs are seen. Both get stronger with energy and both have about
the same amplitude in each band (see Tab. \ref{tab1}).  When \sz$<$0.5
only the higher frequency QPO is seen.  Apart from an overall increase
in amplitude its energy dependence is approximately the same as for
\sz$>$1.0 (see Tab. \ref{tab1}).

The kHz QPOs and HBO and its second harmonic occur simultaneously (see
Fig. {}\ref{powerspectrum}).  The HBO frequency increases from the
left end of the HB onto the upper NB (Fig. \,\ref{QPOs}b).  At
\sz=1.34\pp0.05 it reaches a maximum of 61\pp0.2 Hz, further down the
NB it decreases again.  On the HB the rms amplitude of the fundamental
decreases with \sz; on the NB it is about constant at 1.5--2\%
(Fig.\,\ref{QPOs}d).  Above \sz=1.6 it is not detected with upper
limits of 2.0\% rms.  The rms amplitude of the second harmonic
decreases on the HB with \sz\, (Fig.\,\ref{QPOs}f).  A
detailed analysis of the HBO and N/FBO will be reported by Homan et
al. (1997).

\section{Discussion \label{discussion}}

We detected two simultaneous kHz QPOs in GX 17$+$2. After Sco X-1 (van
der Klis et al 1996a; 1997b) and GX 5$-$1 (van der Klis et al. 1996b)
this is the third Z source displaying kHz QPOs\footnote{Recently, we
also found kHz QPOs in Cygnus X-2 (Wijnands et al. 1997b).}. The QPO
frequency increases down the HB and onto the NB. For atoll sources an
increase in frequency is thought to be due to an increase in
\mdot. Due to the very similar properties of the kHz QPOs in atoll
sources and in GX 17$+$2 our observations confirm that \mdot\, indeed
increases when the source moves down the Z, even when (on the NB) the
count rate decreases, as previously concluded by Hasinger \& van der
Klis (1989).

Although the behavior of the X-ray spectra changes significantly at
\sz=1 when the source enters the NB, and the X-ray flux begins to
decrease with \sz\, there, the frequencies of all QPOs continue to
increase with \sz\, in the same manner as on the HB. The only evidence
that the rapid variability is affected as the source enters the NB is
by the HBO rms amplitude. Decreasing with \sz\, on the HB, the
amplitude is about constant on the NB.

\begin{figure}[t]
\begin{center}
\begin{tabular}{c}
\psfig{figure=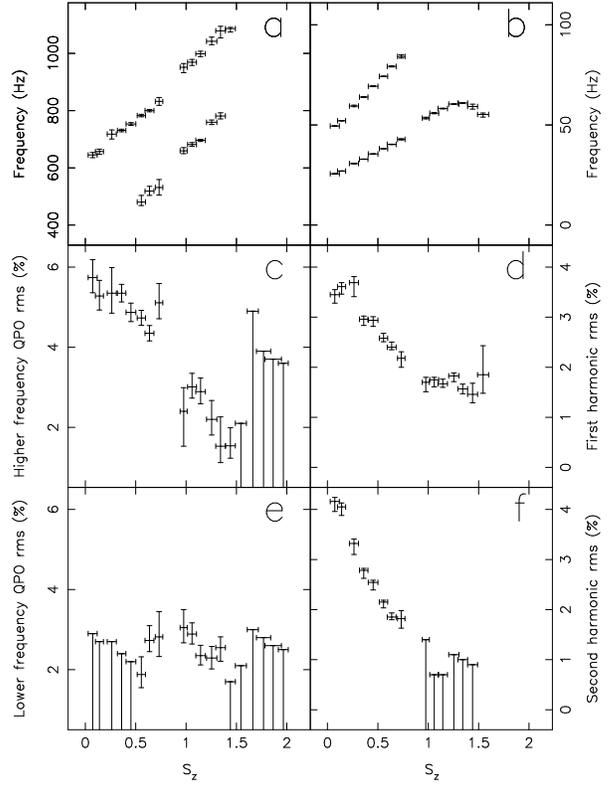,width=8cm}
\end{tabular}
\caption{({\it a}) Frequency of the kHz QPOs, ({\it
b}) frequency of the first and second harmonic of the HBO, ({\it c})
the rms amplitude of the higher frequency kHz QPO, ({\it d}) the
rms amplitude of the first harmonic of the HBO, ({\it e}) the rms
amplitude of the lower frequency kHz QPO, and ({\it f}) the rms
amplitude of the second harmonic of the HBO as a function of
\sz.\label{QPOs}}
\end{center}
\end{figure}

Between \sz=1.4 and 1.5 the HBO frequency decreases.  A similar
decrease in HBO frequency on the NB for GX 17$+$2 was seen in the
``variable-frequency'' QPO on the NB reported by Wijnands et
al. (1996). Although supporting evidence for a decreasing HBO
frequency on the NB was found in Ginga data on Cygnus X-2 (Wijnands et
al. 1997a), it was unclear if the ``variable-frequency'' QPO in
GX\,17$+$2 was indeed the HBO (see Kuulkers et al. 1997). From
Fig. {}\ref{QPOs}b it is now evident that indeed it was, and also that
it was the fundamental and not, as proposed by Kuulkers et al. (1997),
the second harmonic.  The data reported here and by Wijnands et
al. (1996) show that the behavior of the HBO did not change
significantly over one year. In the data presented here the HBO does
not decrease to the same frequency as in 1996; this is due to a lack
of data on the NB.

Beat-frequency models (BFMs) are the best models so far to explain the
kHz QPOs in atoll sources. In GX 17$+$2 kHz QPOs and HBOs are observed
simultaneously, so if the magnetospheric BFM explains the HBO (Alpar
\& Shaham 1985; Lamb et al. 1985) it can not explain the kHz QPOs (the
same is true for Sco X-1; van der Klis et al. 1997b). The sonic-point
BFM for the kHz QPOs in atoll sources (Miller, Lamb, \& Psaltis 1997)
assumes that the highest frequency is due to orbital motion around the
neutron star at the sonic radius, and the lower frequency to its beat
with the neutron star spin frequency. This model predicts a constant
twin peak separation, as observed for atoll sources. According to this
model the spin period of the neutron star in GX 17$+$2 would be
3.41\pp0.03 msec.

\begin{figure}[t]
\begin{center}
\begin{tabular}{c}
\psfig{figure=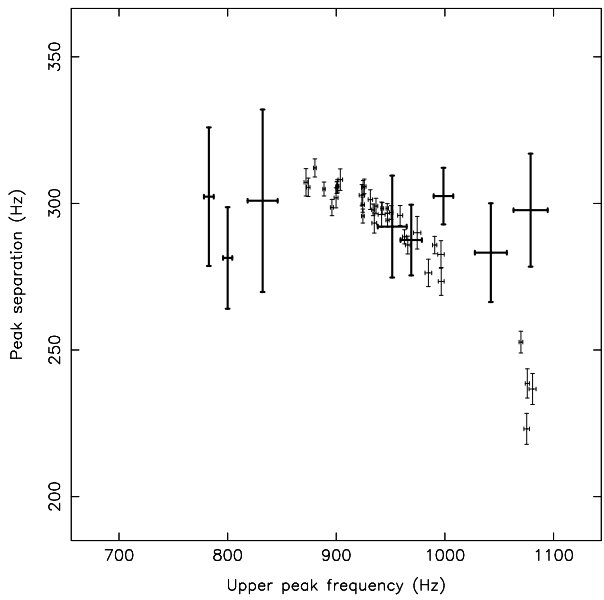,width=8cm}
\end{tabular}
\caption{The kHz peak separation versus the frequency of the higher
frequency QPO. The bold points with the large error bars are for
GX\,17$+$2, the fine points with the small error bars are for Sco X-1
(taken from van der Klis et al. 1997b, their Figure
3a). \label{peak_difference}}
\end{center}
\end{figure}

In Sco X-1 the twin peak separation decreases when the frequencies of the
kHz QPOs increase (van der Klis et al. 1997b). Although the peak
separation in GX 17$+$2 is consistent with being constant, its error
bars are larger than in Sco X-1 and a decrease similar to that in Sco
X-1 can not be excluded (see Fig. \ref{peak_difference}). Note that
the decrease of the peak separation in Sco X-1 was observed when this
source was moving down the NB onto the beginning of the FB. We could
not observe the kHz QPOs this far down the Z in GX 17$+$2.

Another explanation that has been proposed for kHz QPOs is the Photon
Bubble Oscillation (PBO) model of Klein et al. (1996).  These
oscillations have been modeled to occur below 0.4 L$_{\rm Edd}$ and
only weakly (or not at all) above 1.1 L$_{\rm Edd}$ (Klein et
al. 1996; R. Klein, private communication 1997). The amplitudes of the
PBOs decrease and their frequencies increase with \mdot, just as we
observe. However, in calculations so far PBOs near 1200 Hz occur at
0.4 L$_{\rm Edd}$, a factor two lower than the inferred luminosity
when we observed 1200 Hz QPOs. Calculations for luminosities between
0.4 and 1.1 L$_{\rm Edd}$ covering a wider range of parameters are
needed to compare the properties of the PBOs more accurately with the
observed properties of the kHz QPOs in Z sources.

GX 17$+$2 and Sco X-1 have similar timing properties.  The highest
observed frequency of the kHz QPOs in both sources is approximately
1100 Hz, and the peak separations are more or less equal (see
Fig.\,\ref{peak_difference}). However, several aspects of the timing
behavior differ significantly between the two sources. In Sco X-1 the
highest frequency for the kHz QPOs is observed at the beginning of the
FB (van der Klis et al. 1996a), but in GX 17$+$2 this frequency is
already reached on the upper NB. This difference is remarkable,
because both sources are thought to reach L$_{\rm Edd}$ near the NB/FB
vertex (i.e. \sz=2). So, the same frequency is reached in GX\,17$+$2
further from the Eddington critical value than in Sco X-1.

Also, in Sco X-1 the frequency of the HBO does not decrease as \mdot\,
increases (van der Klis et al. 1997b).  Wijnands et al. (1996)
suggested that an increase in the mass flux through the radial inflow
as GX 17$+$2 moves down the NB could produce the decrease of mass flux
through the inner disk needed in the magnetospheric BFM to explain the
observed decrease of the HBO frequency with increasing \mdot. It
remains to be seen if indications can be found for a different radial
inflow behavior between the two sources, as would be required to
explain the difference in the HBO frequency changes with \mdot\,
within this framework.

\acknowledgments

This work was supported in part by the Netherlands Foundation for
Research in Astronomy (ASTRON) grant 781-76-017 and by NSF grant AST
93-15133. B.V. (NAG 5-3340), F.K.L. (NAG 5-2925), J. v. P. (NAG
5-3269, NAG 5-3271), and W.H.G.L. acknowledge support from United
States NASA grants. MM is a fellow of the Consejo Nacional de
Investigaciones Cient\'{\i}ficas y T\'ecnicas de la Rep\'ublica
Argentina.

\clearpage
\clearpage

\clearpage

\begin{deluxetable}{cccc}
\tablecolumns{4}
\tablewidth{0pt}
\tablecaption{kHz QPO rms amplitude versus photon energy$^a$ \label{tab1}}
\tablehead{
\colhead{} & \sz=0.5\pp0.14 &\multicolumn{2}{c}{\sz=1.15\pp0.10}\\
\colhead{} & Upper peak & Upper peak & Lower peak\\
Energy    & rms                 & rms     & rms    \\
(keV)      & (\%)                & (\%)    & (\%)}
\startdata
2--5.0  & $<1.9$     &  $<1.1$    & $<1.2$     \\
5.0--6.4&  3.4\pp0.5 &  2.0\pp0.5 & $<1.9$     \\
6.4--8.6&  3.9\pp0.5 &  2.2\pp0.4 & 1.7\pp0.5  \\
8.6--60 &  7.4\pp0.2 &  3.6\pp0.3 & 4.1\pp0.2  \\
\hline
5.0--60$^b$  & 5.2\pp0.2 &2.1\pp0.2  & 2.3\pp0.2  \\
\enddata
\tablenotetext{a}{All errors correspond to $\Delta\chi^2=1$. The upper
limits correspond to the 95\% confidence level.}
\tablenotetext{b}{Added for comparison with Cygnus X-2 (Wijnands et
al. 1997b).}
\end{deluxetable}

\end{document}